\documentclass[proceedings]{JHEP3} 
\PrHEP{ hep2003}

\usepackage{epsfig,multicol}            

\newbox\mybox
\newcommand\fverb{\setbox\mybox=\hbox\bgroup\verb}
\newcommand\fverbdo{\egroup\medskip\noindent\fbox{\unhbox\mybox}\ }
\newcommand\fverbit{\egroup\item[\fbox{\unhbox\mybox}]}

\title{Detecting  Solar Neutrino Flares and Flavors}

\author{Daniele Fargion\\
    Physics Department, University of Rome P.le A. Moro,2, 00185 ROME,
    ITALY\\
    E-mail: \email{daniele.fargion@roma1.infn.it}}

\abstract{ Intense solar flares originated in sun spots produce
high energy particles (protons, $\alpha$) well observable by
satellites and ground-based detectors. The flare onset produces
signals in different energy bands (radio, X, $gamma$ and
neutrons). The most powerful solar flares as the  ones
   occurred on 23 February  $1956$,  29 September $1989$  and the more recent
   on
     October $28th$, and  the $2nd$, $4th$,  $13th$  of November  $2003$
released in sharp times the largest flare energies (${E}_{FL}
\simeq {10}^{31}\div {10}^{32}$ erg). The high energy solar flare
protons scatter within the solar corona and they must be source of
a prompt neutrino burst through the production of charged pions.
Later on, solar flare particles hitting the atmosphere may
marginally increase the atmospheric neutrino flux. The prompt
solar neutrino flare may be detected in the largest underground
$\nu$ detectors. Our estimate for the October - November 2003
solar flares gives a number of events above the unity. The
electron/muon $\nu$ signals and spectra may reflect the neutrino
flavour mixing.        A surprising $\tau$ appearance may occur
for a  hard (${E}_{{\nu}_{\mu}} \rightarrow
{E}_{{\nu}_{\tau}}\simeq> 4 GeV$) flare spectra. }

\begin{document}


\section{Introduction}           
\label{sect:intro}
The recent peculiar solar flares on October-November $2003$ recalls
  the historical one of February 23th, 1956
\cite{ref0} and the most powerful event occurred on September
$29th$, $1989$ at 11:30 - 12:00 UT \cite{ref1}, \cite{89}. These
events were source of high energetic charged particles whose
observed energies, $E_{p}$, ranged between the values: $15 GeV
\geq E_{p}\geq 100$ MeV, although  even higher proton solar
energies $E_p \geq 500 GeV$ have been reported (see \cite{ref9}).
A large fraction of these {\itshape{primary}} particles, i. e.
solar flare cosmic rays, became a source of both neutrons and
{\itshape{secondary}} kaons, pions $K^{\pm}$, $\pi^{\pm}$ by
their particle-particle spallation on the Sun surface first, and
then on the Earth's atmosphere \cite{ref1}. Consequently,
$\mu^{\pm}$, muonic and electronic neutrinos and anti-neutrinos,
${\nu}_{\mu}$, $\bar{\nu}_{\mu}$, ${\nu}_{e}$, $\bar{\nu}_{e}$,
$\gamma$ rays, are released by the chain reactions $\pi^{\pm}
\rightarrow \mu^{\pm}+\nu_{\mu}(\bar{\nu}_{\mu})$, $\pi^{0}
\rightarrow 2\gamma$, $\mu^{\pm} \rightarrow
e^{\pm}+\nu_{e}(\bar{\nu}_{e})+ \nu_{\mu}(\bar{\nu}_{\mu})$.
There are two different sites for these  decays to occur, and two
corresponding neutrino emissions (see  \cite{ref10}):
\begin{description}
  \item[(1) ] A  brief and sharp solar flare {\itshape{neutrino burst}}, originated within the solar corona;
  \item[(2) ] A diluted and delayed  terrestrial {\itshape{neutrino flux}}, produced by flare particles hitting the
  Earth's   atmosphere.
\end{description}
The first is a prompt  neutrino burst (few seconds/minutes onset)
due to charged particles scattering onto the solar shock waves,
associated with prompt gamma, X, neutron events.  The largest
event which occurred at 19:50 UT  on November 4 2003  was
recorded as an $X28$, the most intense X ray event from our Sun .
The consequent {\itshape{solar}} flare neutrinos reached the Earth
with a well defined directionality and within a narrow time
range. The corresponding average energies $<E_{{\nu}_{e}}>$,
$<E_{{\nu}_{\mu}}>$  are probably larger compared to an event in
Earth's atmosphere since the associated primary particles
(${\pi}^{\pm}$, ${\mu}^{\pm}$) decay in flight at low solar
densities, where they  suffer negligible energy loss:
$<E_{{\nu}_{e}}> $ $\simeq$ $50 MeV$, $<E_{{\nu}_{\mu}}> \simeq$
100 $\div$ 200 MeV.

 The delayed {\itshape{neutrino flux}} originated in the Earth's
atmosphere is due to the arrival of prompt solar charged particles
nearly ten minutes later than onset of  the radio-X emission.
These particles must not be  confused with those at lower energy
originated in more delayed solar winds. Such nearly relativistic
(100 - 1000 MeV) solar flare cosmic rays are charged and bent by
inter-planetary particles and fields. Therefore their arrival and
the corresponding  neutrino production in the Earth's atmosphere
occurs tens minute or even a few hours later than the solar
X-radio sharp event. As a result, their signal is widely spread
and diluted in time. The atmospheric neutrino directions at sea
level, following the cosmic rays arrival maps, are nearly
isotropic or, more precisely they are slightly clustered near the
terrestrial magnetic poles. A large fraction of the energy of the
primary solar flare cosmic rays (such as protons and alpha
particles) is dissipated by ionization in the earth's atmosphere.
Therefore {\itshape{terrestrial}} electronic neutrinos
${\nu}_{e}$, $\bar{\nu}_{e}$, are originated by muons almost at
rest because of the dense terrestrial atmosphere,  and are
leading to a soft terrestrial neutrino flare spectra. Their mean
energy ${E}_{{\nu}\oplus}$ is on average smaller than the original
{\itshape{solar flare}} ones, $<{E}_{{\nu}\odot}>\simeq 100$ MeV,
and their total relic energy ratio (terrestrial neutrino over
solar flare), $\frac{{E}_{{\nu}\oplus}}{{E}_{{\nu}\odot}}\leq
10^{-1}$, is also poor. Because of the quadratic or linear
increase of the cross section with energy, the detection of the
terrestrial neutrino flux is harder than the solar one. Moreover,
the terrestrial diluted neutrino flux may be hidden (excluding few
 cases in present detectors) by the comparable steady atmospheric
neutrino background. For these reasons we may neglect the low
energetic terrestrial {\itshape{neutrino flux}}, even if it is a
well defined  source of secondary neutrinos. Statistically it will
be hard to observe this signal in present Super-Kaimokande (SK)
detector, but they might be observable in the future  larger
underground detectors.

In this paper we  analyse the observable consequences due to the
first prompt solar flare: {\itshape{a solar neutrino burst}}. We
consider two mechanisms to produce neutrinos from proton-proton
scattering: these flare particles may scatter either outward or
inward the solar surface while pointing at the same time to the
(Earth) observer. Because of the very different consequent target
solar atmosphere, the $\pi^{\pm}$, $\mu^{\pm}$, and $\nu_e$,
$\bar{\nu}_e$, $\nu_{\mu}$, $\bar{\nu}_{\mu}$ production is
different. Our estimate of the solar flare {\itshape{neutrino
burst}} is scaled by an integrated flare energy ${E}_{FL}$, which
is assumed to be of the order of
${E}_{FL}\simeq10^{31}\div10^{32}$ erg, by comparison with the
known largest solar flare events as those in $1956$ and $1989$.
The recent  solar flare spectra are unknown but their energies
are extending well above the few GeV threshold necessary to the
pion production. To give a rough idea of the order of magnitude
of a solar neutrino flare on the Earth, we may compare the
$\underline{total}$ flare energy flux ${\Phi}_{FL}$, at the
Sun-Earth distance ${d}_{\odot}$ with the corresponding energy
flux of the well studied supernov\ae~explosion SN1987A, occurred
on the 23rd February  1987 at a distance ${d}_{SN}$:
\begin{equation} \frac{{\Phi}_{FL}}{{\Phi}_{SN}} \simeq \frac{{E}_{FL}}{{E}_{SN}} \left( \frac{{d}_{SN}}{{d}_{\odot}}
\right)^{2}\simeq \frac{1}{30}\left( \frac{E_{FL}}{10^{32}~erg}
\right) \left( \frac{E_{SN}}{3\cdot10^{53}~erg} \right)^{-1}
\end{equation}
The ratio, even if smaller than unity,  remarks the flare energy
relevance. The SN neutrino fluence is probed both experimentally
and theoretically  while the conversion of the solar flare
energy  ($10^{32}~erg$)  in neutrinos has to be probed yet.
However even in a more conservative scenario, where  only a
fraction $\eta <  0.1 $ of the flare energy  is   converted into
neutrinos, the flare energy flux on the Earth is:
\begin{equation}
{\Phi}_{FL}= 3.5\cdot \eta \cdot 10^{4}~erg~cm^{-2}\left(
\frac{{E}_{FL}}{10^{32}~erg} \right)
\end{equation}

We know that Kamiokande  detectors observed $11$ neutrino events
from the 1987A supernov\ae~explosion, while SK, because of its
larger volume, may observe a signal $22$ times as large.
Therefore, even if $\eta < \sim 10\% $ the signal ($\sim 0.8 $)
is near or above unity and it may be reached. Moreover the
expected pion-muon neutrino flare mean energy $<{E}_{{\nu}{FL }}>
\geq kT_{{\nu}{SN}} \simeq 10 $ MeV is much larger  than the
corresponding one for thermal supernov\ae~ neutrinos
(${E}_{{\nu}{FL }}\simeq 0.1-1$ GeV). The consequent $\nu$ $-$
$N$ cross section  increases with the square of the energy,
therefore the event number, to a  first approximation, is larger
(by a factor ten or more) than 10 MeV. Consequently, if a
fraction $\geq 0.001$ of the total flare energy
(${E}_{FL}\simeq{10}^{32}$ erg) was emitted into neutrino relics
via the $p \rightarrow \pi \rightarrow \mu$ chain, their arrival
on Earth might be nearly detectable ($\sim1\div5$ events) and it
would be much worth checking the Super-Kamiokande records at the
corresponding  X-radio precursor flare; for instance the X-ray
maximum on the $28th$ October, during the $2nd-4th$ November
(edge), and the $13th$ November (hidden) solar flare $ 2003$.
Gamma signals  are also expected in similar events:
    there are very limited gamma data (OSSE-EGRET) on solar flares and only a very surprising evidence
    of hard gamma events \cite{Lin} on the July $2002$
    solar flare, where hard X-gamma events have been followed in details by  RHESSI gamma
    detector.
     The associated gamma energy of the $2002$ flare is  smaller  than our neutrino estimate.
      It is worth reminding to the reader that  the
      absence of large gamma solar flares have been used to infer a
      bound on anti-meteorite and  anti-matter presence in our
      solar system and galaxy (see
      \cite{ref4}).
On the other hand, downward gamma and/or  hard $X$ flares while
being closely absorbed might play a key role in enhancing neutrino
signal over the gamma flare. Therefore we consider the total energy
of the flare (both kinetic and X) as
    the main evidence to predict the neutrino signal.

\section{The solar flare energy balance}
\label{sect:the solar}
The energy released during the largest known flares is mainly
under the form of inter-planetary shock waves ${E}_{FL} \geq
{10}^{32}$ erg (see \cite{ref2}) up to  ${E}_{FL} \leq {10}^{33}$
erg, (see \cite{Lin}). A large fraction of energy is found in
optical emission ${E}_{FLop}\simeq 8\cdot{10}^{31}$ erg and in
soft and hard X-rays (by electromagnetic or nuclear
bremsstrahlung), as well as in energetic cosmic rays
($2\div5\cdot{10}^{31}$ erg). The flare particles might be
pointing towards the Earth  or they may be just beyond the solar
disk, as in the September $1989$ event,  located behind the West
limb of the Sun (${105}^{o}$ West) and those on the $4th$ - $6th$
- $13th$ November $2003$. The $1989$ flare was $\underline{first}$
observed at a 8.8 Gigahertz radioburst, (because of the
refractive index of solar atmosphere), at  11:20 U. T. and. Later
it reached a higher (visible) peak in the X-rays domain (see
\cite{ref1}, \cite{ref10}). Is there any {\itshape hidden}
underground flare whose unique trace is in a powerful
(unobserved) neutrino burst?  Secondary gamma rays due to common
neutral pion decay, positron annihilations and neutron capture,
have a very small cross section, thus there must be an
observable  signature on the Sun surface of such a powerful hidden
flare. Nevertheless observed gamma ray flares, are not in favor
of any extreme $E_{FL} \gg 10^{33} erg$ underground flares( see
\cite{89}, \cite{Lin}). It must be kept in mind that the rarest
event on February $'56$ was not observed at gamma wavelengths,
because of the absence of such satellite detectors at the time,
while the $Sept. 29th 1989 $ event was not detected in gamma rays
because it occurred on the opposite solar side. On the other hand
lower powerful solar flares as that of the $4th June 1991$ have
been studied in all radio-X-gamma energy up to tens MeV energy by
the OSSE detector of the CGRO satellite. Therefore there are no
direct bounds on a larger {\itshape hidden} underground flare.
One may suspect that  a too large solar flare event in its hidden
side should be reflected somehow into an electromagnetic cascade
which may influence the continuous solar energy spectrum
(${E}_{\odot}\simeq 3.84 \cdot{10}^{33}$ erg $s^{-1}$), even in
the observable side. Moreover recent accurate heliosismography
might be able to reveal any extreme hidden flare energy. We may
therefore restrict our most powerful solar flare energy in the
range:
\begin{equation}
{10}^{33}~erg \sim\geq {E}_{FL}\geq {10}^{31}~erg
\end{equation}
keeping the lowest value  as the  flare energy threshold.
\subsection{The proton-proton pion production in solar flare}
\label{sect:the}
The kaon-pion-muon chain reactions and their consequent neutrino
relics spectrum in solar atmosphere may be evaluated in detail if
the energetic particle (protons, alpha nuclei, ...) energy spectra
is known, as well as the solar density and magnetic configuration.
 Indeed magnetic screening may reduce high energy particle scattering
 in the solar flare regions. Successful description for terrestrial atmospheric
 neutrinos, and their primary relic of cosmic rays, has been obtained. Our
approach, ignoring the exact spectrum for protons in recent solar
flare and the detailed magnetic configuration, will force us to
consider only averaged values, neglecting  the (higher energetic)
Kaon production. In order to find the interaction probability for
an energetic proton (${E}_{p} \simeq 2$ GeV) to scatter
inelastically with a target proton at rest in solar atmosphere, we
must assume an exponential solar density function following the
well known solar density models (reference \cite{ref5}).
\begin{equation}
{n}_{\odot}={N}_{0}{e^{\frac{-h}{h_{0}}}};~
{N}_{0}=2.26\cdot{10}^{17}~{cm}^{-3},
~{h}_{0}=1.16\cdot{10}^{7}~cm
\end{equation}
where $h_{0}$ is the photosphere height where flare occurs.
\subsection{Protons interactions up-going vertically in solar flare}
\label{sect:upward}
The inelastic proton-proton cross section for energetic particles
(${E}_{p}>2$ GeV) is nearly constant: ${\sigma}_{pp}(E>2~GeV)\simeq
4\cdot{10}^{-26}~{cm}^{2}$. Therefore the scattering probability
${P}_{up}$ for an orthogonal upward energetic proton ${p}_{E}$, to
produce pions (or kaons) via nuclear reactions is:
\begin{equation}
{P}_{up}={1-{e^{-\int^{\infty}_{h_{0}=0}
{\sigma}_{pp}n_{\odot}dh}}}\simeq 0.1
\end{equation}
A terrestrial Observatory whose line of sight includes the solar
flare would observe only 10\% (or much less, if, as it is well
possible ${h}_{0}> 10^7 cm$) of the primordial proton flare number,
converted into pions and relic muons, neutrinos and
electron-positron pairs. Moreover, because of the kinematics, only
a fraction smaller than 1/2 of the energetic proton will be
released to pions (or kaons) formation. In the simplest approach,
the main source of  pion production is $p+p\rightarrow
{{\Delta}^{++}}n\rightarrow p{{\pi}^{+}}n$; $p+p\rightarrow
{{{\Delta}^{+}}p}^{\nearrow^{p+p+{{\pi}^{0}}}}_{\searrow_{p+n+{\pi}^{+}}}$
at the center of mass of the resonance ${\Delta}$ (whose mass value
is ${m}_{\Delta}=1232$ MeV). The ratio ${R}_{{\pi}{p}}$ between the
pion to the proton energy is:
\begin{equation}
 {R}_{{\pi} p}=
\frac{{E}_{\pi}}{{E}_{p}}=\frac{{{{m}_{\Delta}}^{2}}+{{{m}_{\pi}}^{2}}
-{{{m}_{p}}^{2}}}{{{{m}_{\Delta}}^{2}}+{{{m}_{p}}^{2}}-{{{m}_{\pi}}^{2}}}=0.276
\end{equation}
Therefore the total pion flare energy due to upward proton is:
\begin{equation}
{E}_{{\pi}_{FL}}=P{R}_{{\pi}p}{E}_{FL}=2.76\cdot{10}^{-2}{E}_{FL}
\end{equation}
Because of the isotopic spin, the probability to form a charged
pion over a neutral one in the reactions above: $p+p\rightarrow
p+n+{\pi}^{+}$, $p+p\rightarrow p+p+{\pi}^{0}$, is given by the
Clebsh Gordon coefficients, (3/4), and by the positive-negative
ratio (1/2):
\begin{equation} {C}_{\frac{{\pi}^{-}}{{\pi}^{0}}}\simeq
{C}_{\frac{{\pi}^{+}}{{\pi}^{0}}}\simeq \frac{3}{8}
\end{equation}
The ratio of the neutrino and muon energy in
pion decay is also a small adimensional  fraction
${R}_{{\nu}_{\mu}{\mu}}$
\begin{equation}
 {R}_{{\nu}_{\mu}{\mu}} =
\frac{{E}_{{\nu}_{\mu}}}{{E}_{\mu}}=\frac{{{m}_{\pi}}^{2}-{{m}_{\mu}}^{2}}
{{{m}_{\pi}}^{2}+{{m}_{\mu}}^{2}}=0.271
\end{equation}
To a first approximation one may assume that the total pion energy
is equally distributed in all its final remnants:
($\bar{\nu}_{\mu}$, ${e}^{+}$, ${\nu}_{e}$, ${\nu}_{\mu}$) or
(${\nu}_{\mu}$, ${e}^{-}$, $\bar{\nu}_{e}$, $\bar{\nu}_{\mu}$):
\begin{equation}
 \frac{{E}_{\bar{\nu}_{\mu}}}{2}\simeq
\frac{{E}_{{\nu}_{\mu}}}{2}\simeq {E}_{{\nu}_{e}}\simeq
{E}_{\bar{\nu}_{e}}\simeq \frac{1}{4}{E}_{{\pi}^{+}}
\end{equation}
The correct averaged energy (by Michell parameters) for neutrino
decay ${\mu}^{\pm}$ at rest are:
${E}_{\bar{\nu}_{e}}={E}_{{\nu}_{e}}=\frac{3}{10}{m}_{\mu}\simeq
\frac{1}{4}{m}_{\pi};$
$${E}_{\bar{\nu}_{\mu}}={E}_{{\nu}_{\mu}}\simeq \frac{9}{20}{m}_{\mu}\simeq\frac{1}{3}{m}_{\pi}$$

Similar reactions (at lower probability) may also occur by
proton-alfa scattering leading to: $p+n\rightarrow
{{\Delta}^{+}}n\rightarrow n{{\pi}^{+}}n$; $p+n\rightarrow
{{{\Delta}^{o}}p}^{\nearrow^{p+p+{{\pi}^{-}}}}_{\searrow_{p+n+{\pi}^{o}}}$.
Here we neglect their additional role due to the flavor mixing and
the dominance of previous reactions at soft flare spectra.

Therefore  ${E}_{{\nu}_{\mu}}>{E}_{{\nu}_{e}}$: however muon
neutrino from pion ${\pi}^{\pm}$ decays have a much lower mean
energy and the combined result in eq.($2.8$) is a good
approximation. We must consider also  the flavour mixing (in vacuum
) that leads to an averaged neutrino energy along its path.
To a first approximation the oscillation will lead to a $50\%$
decrease in the muon component and it will make the electron
neutrino component harder. We take into account this flavor mixing
by a conversion term $\eta_{\mu} =\simeq \frac{1}{2}$, re-scaling
the final muon neutrino signal and increasing the electron spectra
component.
 Because in ${\pi}$-${\mu}$ decay the ${\mu}$ neutrinos relic
are twice the electron ones, the anti-electron neutrino flare
energy is, at the birth place on Sun:
\begin{equation}
 {E}_{\bar{\nu}_{e}FL}\simeq
{E}_{{{\nu}_{e}}FL}  \simeq\frac{{E}_{{\nu}_{{\mu}} FL}}{2} \simeq
{P}_{up}{R}_{{\pi} p}{C}_{\frac{{\pi}^{-}}{{\pi}^{0}}}{E}_{FL}
\simeq 2.6\cdot{10}^{28}\left( \frac{{E}_{FL}}{{10}^{31}~erg}
\right)~erg.
\end{equation}
The corresponding neutrino flare energy and number fluxes at sea
level are:
\begin{equation}
{\Phi}_{\bar{\nu}_{e}FL} \simeq 9.15 \left(
\frac{{E}_{FL}}{{10}^{31}~erg} \right)~erg~{cm}^{-2}
\end{equation}
\begin{equation}
{N}_{{\nu}_{e}} \simeq {N}_{\bar{\nu}_{e}} \simeq
5.7\cdot{10}^{4}\left( \frac{{E}_{FL}}{{10}^{31}~erg} \right)
\left( \frac{<{E}_{\bar{\nu}_{e}}>}{100~MeV} \right)^{-1} cm^{-2}
\end{equation}

This neutrino number is larger but comparable with a different
value calculated elsewhere (reference \cite{ref6}). The flux
energy in eq.(2.10) is nearly $4000$ times smaller than the
energy flux in eq.($1.2$) and, as we shall see, it may be nearly
observable by present detectors. This flux at $GeV$ energy may
correspond approximately to a quarter of a day atmospheric
neutrino integral fluence (for each flavor specie). Therefore it
may lead to just a half  of an event as occurred on  the $28th$
October $2003$. The largest neutron and gamma flare energies
should be (and indeed are) comparable or even much larger
($February 1956$) than the upward neutrino flux energy in
eq.($2.10$). The exceptional solar flare on $Sept. 29th, 1989$ as
well as the most recent on $2nd -4th November 2003 $ took place
in the nearly hidden disk side and we may look now for their
horizontal or down-ward secondary neutrinos. Their scattering are
more effective and lead to a larger pion production. The
processes we describe here are analogous to those considered for
horizontal and upward neutrino induced air-showers inside the
Earth Crust (see  \cite{Fargion 2002}) and nearly ultra high
horizontal showers (detectable by EUSO). The solar neutrino flare
production is enhanced by a higher solar gas density where the
flare beam occurs. Moreover a beamed X-flare may suggest  a
corresponding beamed pion shower whose mild beaming naturally
increases the neutrino signal. Most of the downward neutrino
signal  discussed in the next sections is generated at mild
relativistic regime as well as their pion and muon secondaries.
Therefore a mild outward neutrino burst  may be expected outside
the sun surface even when the flare is hitting downward (with a
little anisotropy suppression) pointing back towards the Earth.

\subsection{ Proton interactions for vertical down-going  solar flare}
 \label{sect:downward}
High energetic protons flying downward (or horizontally) to the Sun
center  are crossing larger (and deeper) solar densities and their
 interaction probability ${P}_{d}$, is larger than the previous
one (${P}_{up}$) (see section \S 2.2). The proton energy losses due
to ionization, at the atmospheric solar densities where most of the
$p - p $ scattering take place, are low  respect to the nuclear
ones and most of the proton flare energy is converted into
pion-Kaon nuclear productions with few losses.

If the proton direction is tangent to the sun surface or if the
protons are travelling  downward towards the solar core, the
interaction probability is even larger than one. Consequent
unstable and short lived pions of few GeV will decay in flight
because of repetitive nuclear reactions at those solar atmosphere
densities are rare; the pion number density ${n}_{\pi}$ is
described by the following equation:
\begin{eqnarray}
\frac{{dn}_{\pi}}{dt}&=&\int\int[{{\frac{{d}^{2}{{n}_{pE}}}{dEd{\Omega}}}{n}_{pT}{\sigma}_{pp}{v}_{pE}}-
{{\frac{{d}^{2}{{n}_{pE}}}{dEd{\Omega}}}{n}_{\pi}{\sigma}_{p{\pi}}{v}_{\pi}}-
{{\frac{{d}^{2}{{n}_{\pi}}}{dEd{\Omega}}}{n}_{\pi}{\sigma}_{{\pi}{\pi}}{v}_{\pi}]{dEd{\Omega}}}+\nonumber\\
&-&\int^{\infty}_{{m}_{\pi}}\frac{{d}^{2}{{n}_{\pi}}}{dE}{\Gamma}_{\pi}(\frac{{m}_{\pi}}{{E}_{\pi}}){dE}_{\pi};
\end{eqnarray}

where $n_{pE}$, $n_{pT}$, $n_{\pi}$ are the number density of the
flare energetic and target protons, $\sigma_{pp}(E)$,
$\sigma_{p\pi}(E)$, $\sigma_{\pi\pi}(E)$, are the p-p, p-$\pi$,
$\pi$-$\pi$ cross sections. The velocities $v_{pE}$, $v_{\pi}$ are
near the velocity of light and $\Gamma_{\pi}=3.8\cdot10^{7}s^{-1}$.
The last term in eq.$(2.12)$, due to the relativistic pion decay,
at solar densities as in eq.$(2.2)$ and at an energy
$E_{\pi}\simeq$ GeV, is nearly six order of magnitude larger than
all other terms, therefore the pion number density $n_{\pi}$ should
never exceed the corresponding proton number density ${n_{pE}}$.
However the integral number of all pion stable relics
($\bar{\nu}_{\mu}$, $\nu_{\mu}$, $\nu_{e}$, $\bar{\nu}_{e}$,
$e^{-}$) may exceed, in principle, the corresponding number of
proton flare, because each proton may be a source of more than one
pion chain. The proton number density below the photosphere ($h<0$)
is described by a polytropic solution, but it may also be
approximated by a natural extrapolation of the Eq.$(2.2)$ with a
negative height $h$.  It is easy to show that the interaction
probability for a relativistic proton ($E_{pE}>> $ GeV) reaches
unity at depth $h=-278$ Km which is the interaction length. At the
corresponding density ($n\odot\sim2.2\cdot10^{18}cm^{-3}$) the
proton ionization losses, between any pair of nuclear reactions are
negligible (few percent).  Unstable relic pions decay (almost)
freely after a length
$L_{\pi}\simeq(\frac{E_{\pi}}{m_{\pi}}){\Gamma_{\pi}}^{-1}C\simeq
7.8\cdot10^{2}(\frac{E_{\pi}}{m_{\pi}})$ cm. The secondary muons
$\mu$ do not loose much of their energy ($\leq 1\%$) in ionization,
($E_{\mu}\leq 0.1-1$ GeV) during their nearly free decay: the muon
flight distance is
$L_{\mu}=\frac{E_{\mu}}{m_{\mu}}{\Gamma_{\mu}}^{-1}C=
6.58\cdot10^{4}(\frac{E_{\mu}}{m_{\mu}})$ cm, and the ionization
losses are:
$\frac{dE_{\pi}}{dx}\simeq\frac{dE_{\mu}}{dx}\simeq10^{-5}~MeV~
cm^{-1}$. In conclusion most of the solar flare energy will
contribute to downward muon energy with an efficiency $\eta$ near
unity. At deeper regions, near $h<-700$ Km where the solar density
is ${n_{\odot}}\geq 10^{20}~cm^{-3}$, a GeV-muon will dissipate
most of its energy in ionization before decaying.
 In that case the energy ratio between muon relics and primary
protons is much smaller than unity.  Only a small fraction of
protons will reach by a random walk such deeper regions and we may
conclude that, in general, the flare energy relations are:
$$E_{{\pi}FL}\equiv{\eta}E_{FL}\leq E_{FL}$$
\begin{equation}
E_{\bar{\nu}_{e}FL}\simeq
E_{{\nu_{e}FL}}\simeq\frac{E_{{\nu_{\mu}FL}}}{2}\simeq
\frac{E_{\bar{\nu}_{\mu}FL}}{2} \simeq
\frac{{\eta}{C_{\frac{\pi^{-}}{\pi_{0}}}}{E_{FL}}}{4} \simeq
9.4\cdot10^{30}\eta \left( \frac{E_{FL}}{10^{32}~erg} \right)~erg
\end{equation}
This result is nearly 36 times larger than the corresponding one
for {\itshape up-ward} neutrinos in Eq.(2.8).  Terrestrial
neutrino relics from cosmic rays produced by pion chain
reactions, lead to a predicted and observed asymmetry (see
\cite{ref7}) between $\bar{\nu}_{e}$, $\nu_{e}$, due to the
positive proton charge predominance both in target and incident
beam:
$$\frac{N_{\nu_{e}}}{N_{\bar{\nu}_{e}}}=\frac{N_{\mu^{+}}}{N_{\mu^{-}}}\simeq1.2$$  at
energies $10$ GeV $>E_{\nu}>100$ MeV. Therefore the energy
component of the observable flare should be marginally reduced in
eq.$(2.13)$, even assuming a low flare output ($10^{31} erg$):
\begin{equation}
 E_{\bar{\nu}_{e}}\simeq 7.8\cdot10^{-2}{\eta}E_{FL}=
 7.8\cdot10^{29}{\eta} \left( \frac{E_{FL}}{{10}^{31}~erg} \right)~erg
\end{equation}

\section{Solar neutrino flare  events in SK-II}
\label{sect:detectable}
We cannot say much about the solar flare neutrino spectrum because
of our ignorance on the recent primordial proton flare spectra. The
solar flare are usually very soft. We may expect a power spectrum
with an exponent equal or larger than the cosmic ray proton
spectrum. Therefore we consider here only averaged neutrino energy
$<E_{\nu}>$ at lowest energies (below near GeV) and we scale the
result above, Eq.$(2.14)$, for the anti-neutrino numbers at sea
level:
\begin{equation}
<N_{\bar{\nu}_{e}}> \simeq 1.72\cdot{10}^{6}{\eta} \left(
\frac{<E_{\bar{\nu}_{e}}>}{100~MeV} \right)^{-1} \left(
\frac{E_{FL}}{{10}^{31}~erg} \right)~{cm}^{-2}
\end{equation}
\begin{equation}
<N_{\bar{\nu}_{\mu}}> \simeq 4.12\cdot10^{6}{\eta} \left(
\frac{<E_{\bar{\nu}_{\mu}}>}{100~MeV} \right)^{-1} \left(
\frac{E_{FL}}{10^{31}~erg} \right)~{cm}^{-2}
\end{equation}
We now consider the neutrino events due to these number fluxes at
Super-Kamiokande II; other detectors as SNO (and AMANDA if the
spectra was extremely hard) might also record a few events but at
much lower rate. The observable neutrino events, due to inverse
beta decay ($\bar{\nu}_{e}+p \rightarrow n+e^{+}$;
$\bar{\nu}_{\mu}+p \rightarrow \mu^{+}+n$), at Super-Kamiokande
detectors are:
\begin{equation}
N_{ev}=\sum_{i}\int\frac{dN_{\bar{\nu}_{i}}}{dE_{i}}{{\sigma}_{\bar{\nu}_{i}p}}(E_{{\nu}_{i}})N_{p_{SK}}dE_{i}
\end{equation}
$ i=e,{\mu}$.
A comparable  neutrino events, due to stimulated beta decay
(${\nu}_{e}+ n \rightarrow p+e^{-}$; ${\nu}_{\mu}+ n \rightarrow
\mu^{-}+ p $), must also take place (see an updated reference
\cite{Bodek}).
 We may approximate this number with an averaged one
due to an effective neutrino energy $\bar{E}_{\nu}$:
\begin{equation}
{N}_{ev}=\sum_{i}<N_{{\bar{\nu}}_{i}}>{\sigma}_{\bar{\nu}_{i}p}(\bar{E}_{{\nu}_{i}}){N}_{p_{SK}}
\end{equation}
Where ${N}_{p_{SK}}$ is the proton number in the Super-Kamiokande
detector ${N}_{p_{SK}}=\frac{{N}_{p}}{{N}_{{H}_{2}O}}N_{nucl}$;
$N_{nucl}=22KT\cdot N_{A}=3.33\cdot10^{34}$; $ \frac{{N}_{p
}}{{N}_{{H}_{2}O}}= \frac{8}{18}$; ${N}_{p_{SK}}=7.38\cdot
10^{33}$. The cross section is an   elaborated analytical formula
(see Strumia et all. $2003$ \cite{strumia}).
 This expression, in agreement with a full detailed result
within few thousandths for $ E_{\nu} \leq 300 MeV $, is
\begin{equation}
\label{naive+} \sigma(\bar\nu_e p)\approx
10^{-43}\,\mbox{cm}^2~p_e E_e ~E_\nu^{-0.07056+0.02018\ln
E_\nu-0.001953\ln^3 E_\nu}, \qquad E_e = E_\nu - \Delta
\end{equation}
 Where $\Delta = m_n - m_p$; $E_e$ is the energy of the escaping
 electron. In a simpler low-energy approximation (see Bemporad et all.,
$2002$ e.g.  \cite{bemporad})
\begin{equation}
\label{simple} \sigma\approx 9.52\, \times 10^{-44} \frac{p_e
E_e}{\hbox{MeV}^2}\ \mbox{cm}^2,\qquad E_e=E_\nu \pm \Delta\hbox{
for }\bar{\nu}_e\hbox{ and }\nu_e,
\end{equation}
  In a more direct form (see  \cite{ref10}),
  at low energy ($10 MeV \leq {E}_{\bar{\nu}_{e}} \leq GeV $)
\begin{equation}
 {\sigma}_{\bar{\nu}_{e}p} \simeq
7.5\cdot10^{-44} \left( \frac{{E}_{\bar{\nu}_{e}}}{MeV}
\right)^{2}cm^{2}
\end{equation}
The expected neutrino event,  during the flare may increase twofold
as we mentioned above because of both  a solar burst and a later
diluted terrestrial flux; for the terrestrial neutrino flux (during
the recent $28/29 Oct. 2003$ flare) due to solar protons hitting
the atmosphere we expect at least :
\begin{equation}
{N}_{ev} \cong 1.7\cdot10^{6}\cdot 7.38\cdot 10^{33}\cdot 6\cdot
10^{-40}\simeq 7.5 \cdot \eta \left(  \frac{E_{FL}}{10^{31} \,
erg} \right)
\end{equation}
 These events in the {\em terrestrial flux} should  almost double  the common atmospheric
neutrino flux background ($5.8$ event a day). For the prompt
neutrino solar burst in the Sun we expect (if occurred in the
hidden or horizontal solar disk) a similar number in a very narrow
time window. Naturally this result might be
 too optimistic. In order to obtain a more severe
result we now tune our expectation with the event number due to the
well known supernov\ae~SN1987A where we know (or we hope to know)
the primordial neutrino energy: $\sum{E}_{{\nu}_{SN}} \simeq
3\cdot10^{53}$ erg and $\bar{E}_{\bar{\nu}_{e}}\simeq 10 $ MeV. We
know (by cosmology and $Z_{0}$ width decay in LEP) that the
possible neutrino flavours states are ${N}_{F}=6$ (${\nu}_{e}$,
$\bar{\nu}_{e}$, $\nu_{\mu}$, $\bar{\nu}_{\mu}$, $\nu_{\tau}$,
$\bar{\nu}_{\tau}$). The Earth-SN1987A distance
$d_{SN}=1.5\cdot10^{23}$ cm leads to:
\begin{equation}
 {N_{ev}}_{\bar{\nu}_{e}}
=\frac{{N}_{{\nu}_{SN}}}{{N}_{F}}{\sigma}_{\bar{\nu}_{e}p}
(\bar{E}_{{\nu}_{SN}}){{N}_{pSK}}= 11 \left(
\frac{E_{SN}}{3\cdot10^{53}~erg} \right) \left(
\frac{\bar{E}_{\nu}}{10~MeV} \right)
\end{equation}
It should be noted that the quadratic energy $\bar{E}_{\nu}$
dependence of the cross section $\sigma_{\bar{\nu}_{e}p}$ and the
inverse energy $\bar{E}_{\nu}$ relation of the neutrino flux number
leads to the linear dependence in Eq.$(3.7)$.  However, the inverse
beta decay processes increases linearly with energy $\bar{E}_{\nu}$
up to a value smaller than $m_{p}\sim$ GeV, above which the cross
section in eq.(3.4) becomes  flat and only at higher energies it
grows linearly with the energy: $\bar{\nu}_{e}+p\rightarrow
e^{+}+n;~~~\bar{\nu}_{\mu}+p\rightarrow \mu^{+}+n;$
${\nu}_{e}+n\rightarrow e^{-}+p;~~~{\nu}_{\mu}+n\rightarrow
\mu^{-}+n;$
\begin{equation}
\sigma_{\bar{\nu}_{e}{p}}\simeq6.2\cdot10^{-39}~cm^{2}
\left(\frac{\bar{E}_{\bar{\nu}_{e}}}{GeV} \right);~~~~
\sigma_{{\nu}_{e}{n}} \simeq 3.5\cdot10^{-39}~cm^{2}\left(
\frac{\bar{E}_{\nu_{e}}}{GeV} \right)cm^{2}
\end{equation}
The formulas above are  approximations only within an  energy
window
$E_{\nu_{\mu}},E_{\bar{\nu}_{\mu}},E_{\nu_{e}},E_{\bar{\nu}_{e}}\simeq
100-1000$ MeV.  As we shall see, we may neglect the prompt
neutrino-electron scattering processes due to charged or neutral
current cross sections:
\begin{eqnarray}
\sigma_{{\nu}_{e}e}\simeq
9\cdot10^{-45}\left(\frac{\bar{E}_{\nu}}{MeV}\right)cm^{2};~
\sigma_{{\nu}_{\mu}e}\simeq1.45\cdot10^{-45}\left(\frac{\bar{E}_{\nu}}{MeV}\right)cm^{2}\nonumber\\
\sigma_{\bar{\nu}_{e}e}\simeq3.7\cdot10^{-45}\left(\frac{\bar{E}_{\nu}}{MeV}\right)cm^{2};~
\sigma_{\bar{\nu}_{\mu}e}\simeq1.24\cdot10^{-45}\left(\frac{E_{\nu}}{MeV}\right)cm^{2}
\end{eqnarray}
Indeed these values are nearly $100$ times smaller (at
$\bar{E}_{\nu}\sim100$ MeV) than the corresponding nuclear ones
in eq.$(3.7)$ and eq.$(3.10)$.  We consider the neutrino flare
signals at Super-Kamiokande due to either $\bar{\nu}_{e}+p
\rightarrow n+e^{+}$ and $\bar{\nu}_{\mu}+p \rightarrow
\mu^{+}+n$, ${\nu}_{e}+n\rightarrow
e^{-}+p;~~~{\nu}_{\mu}+n\rightarrow \mu^{-}+n;$
keeping in mind, for the latter, the threshold energy
  ${E}_{{\nu}_{\mu}}$,${E}_{\bar{\nu}_{\mu}}>113$ MeV. We may summarize from eq.(3.1)
the expectation event numbers at Super-Kamiokande as follows:
$ {N_{ev}}_{\bar{\nu}_{e}} \simeq
0.63{\eta}(\frac{\bar{E}_{\bar{\nu}_{e}}}{35
~MeV})(\frac{E_{FL}}{10^{31}~erg});~\bar{E}_{\bar{\nu}_{e}}\leq
100 ~MeV $;
$ {N_{ev}}_{\bar{\nu}_{e}} \simeq
1.58{\eta}(\frac{E_{FL}}{10^{31}~erg});~
\bar{E}_{\bar{\nu}_{e}}\geq100-1000 ~MeV $;
$ {{N}_{ev}}_{\bar{\nu}_{\mu}} \simeq
3.58{\eta}(\frac{{E}_{FL}}{{10}^{31}~erg});~
\bar{E}_{\bar{\nu}_{\mu}}\geq 200-1000 ~MeV $;
where ${\eta}\leq1$. The neutrino events in Super-Kamiokande may
be also recorded as stimulated beta decay on oxygen nuclei .
Indeed such reactions exhibit two possible channels: $\nu_{e}+O
\rightarrow F+ e^{-},~\bar{\nu}_{e}~O \rightarrow N+e^{+}$; they
have been analyzed by W. C. Haxton $1987$ \cite{ref8}).  For this
reason our preliminary estimate is just a lower bound  for any
high energetic ($E_{{\nu}_{e}}>100$ MeV) neutrino spectrum.
\subsection{The surprising role of Solar Neutrino  Flavor mixing: the $\tau$ appearence }
  The Earth-Sun distance $D_{\oplus\odot}$ is large enough to guarantee a complete
  flavor mixing even for hundred MeV or GeV neutrino energies.
   Indeed the oscillation distance in vacuum is
   $
 L_{\nu_{\mu}-\nu_{\tau}}=2.48 \cdot10^{9} \,cm \left(
 \frac{E_{\nu}}{10^{9}\,eV} \right) \left( \frac{\Delta m_{ij}^2
 }{(10^{-2} \,eV)^2} \right)^{-1} \ll D_{\oplus\odot}=1.5\cdot
 10^{13}cm.
$
   The consequent flavor mixing will increase the average energy of the
 anti neutrino  electron component respect to the one at its birth.
 This will also increase the neutrino electron component while it will reduce
 the corresponding muon component leading to :

 ${\frac{\eta_{\mu}}{\eta_{e}}\simeq \frac{1}{2}}$  and  to
      ${N}_{{ev}_{\bar{\nu}_{\mu}}}\simeq {N}_{{ev}_{\bar{\nu}_{e}}}
       \simeq 2 (\frac{<{E}_{{\nu}_{\mu}}>}{200~MeV}(\frac{<{E}_{FL}>}{{10}^{31}~erg})$ ;
     ${N}_{{ev}_{{\nu}_{\mu}}}\simeq {N}_{{ev}_{{\nu}_{e}}} $
       as well as a comparable, ${\nu}_{e}$, ${\nu}_{\mu}$, $\bar{\nu}_{e}$, $\bar{\nu}_{\mu}$
        energy fluence and  spectra.
        At energies above the $\tau$ threshold energy ${E}_{{\nu}_{\mu}} \geq 3.46 $ GeV a
         surprising $\tau$ appearance may occur: this requires
          a hard (${E}_{{\nu}_{\mu}} \rightarrow {E}_{{\nu}_{\tau}}$$\simeq  4 GeV$) flare spectra.

 Any positive evidence for such events will mark a new
road to Neutrino Astrophysics, to be complementary to lower
neutrino energy from Sun and Supernov\ae. New larger generations of
neutrino detectors will be more sensitive to such less powerful,
but more frequent and energetic solar flares, than to the rarest
extragalactic supernov\ae~ events (as the one from Andromeda).

\EPSFIGURE[ht]{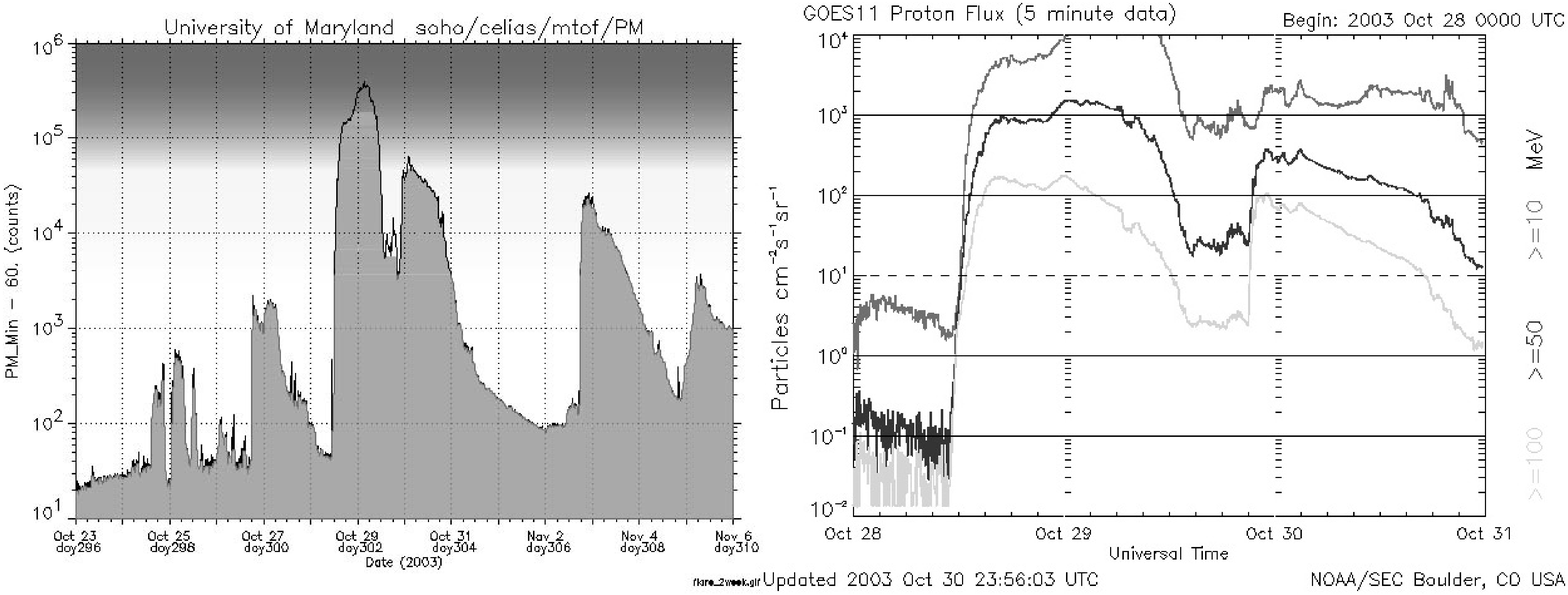,width=13cm}{Proton
solar flare flux (left: all the events during 23 October-6
November 2003 at lowest energies);
    (right:  detail of part (10-100 Mev) energy the spectra);
  the data are respectively from SOHO and GOES11 satellite experiments.}

\EPSFIGURE[ht]{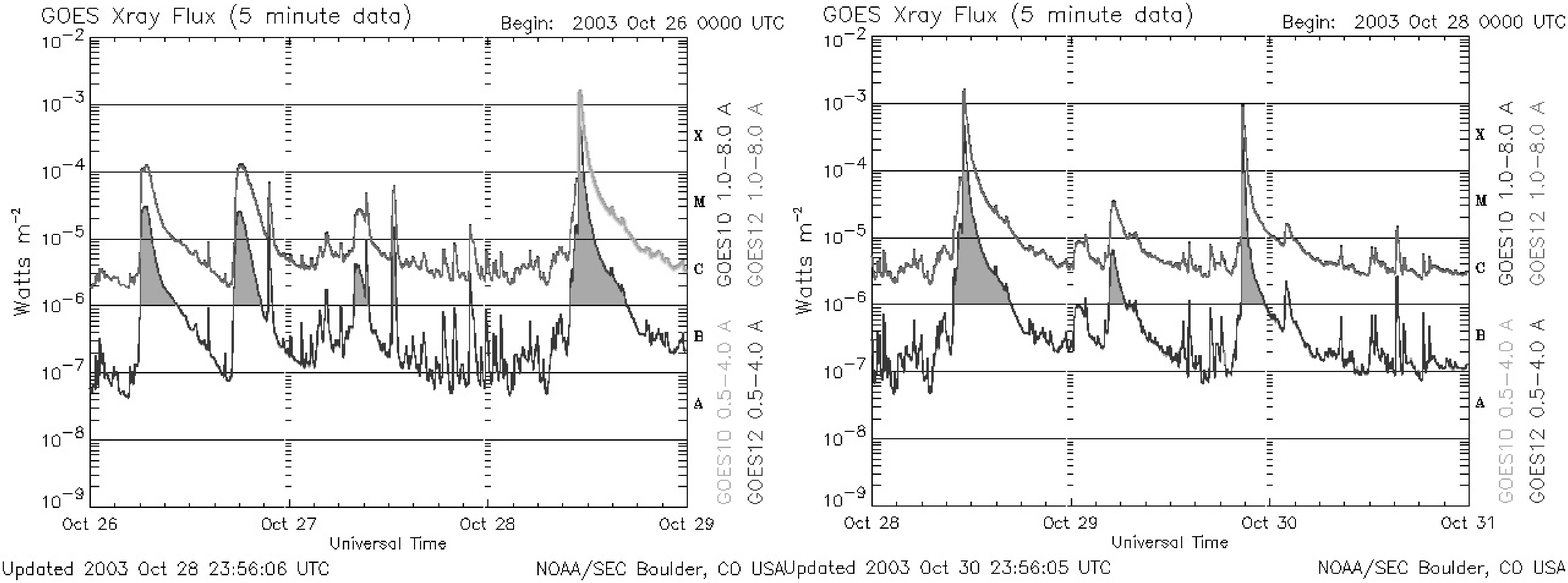,width=13cm}{X Solar Flare
on-set on 26-29th October 2003 by GOES satellite: readable are
the X-peak outburst.}

\EPSFIGURE[ht]{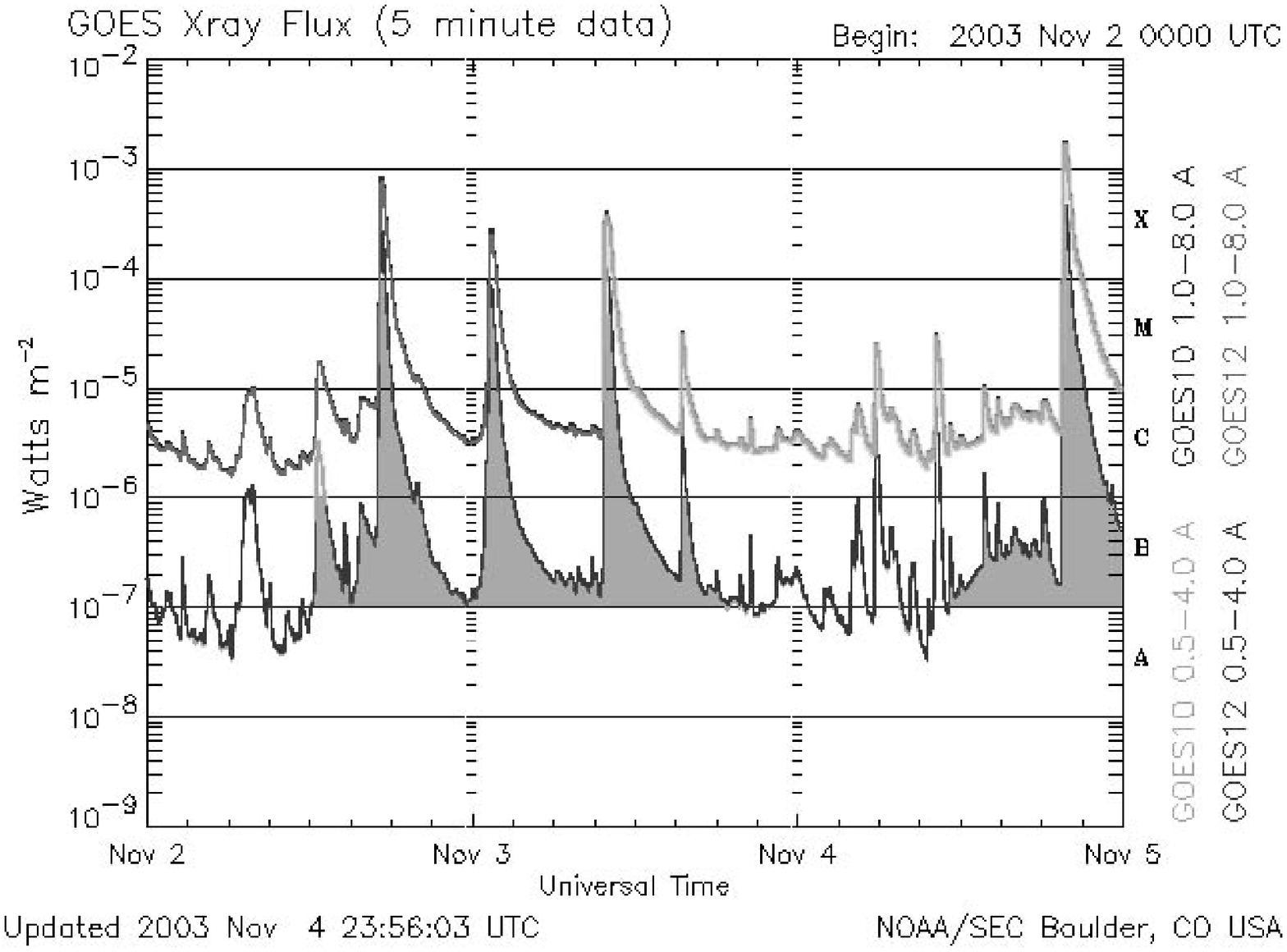,width=10cm}{X Solar Flare on-set
2nd-4th November 2003 by GOES satellite: readable are the X-peak
outburst.}

\EPSFIGURE[ht]{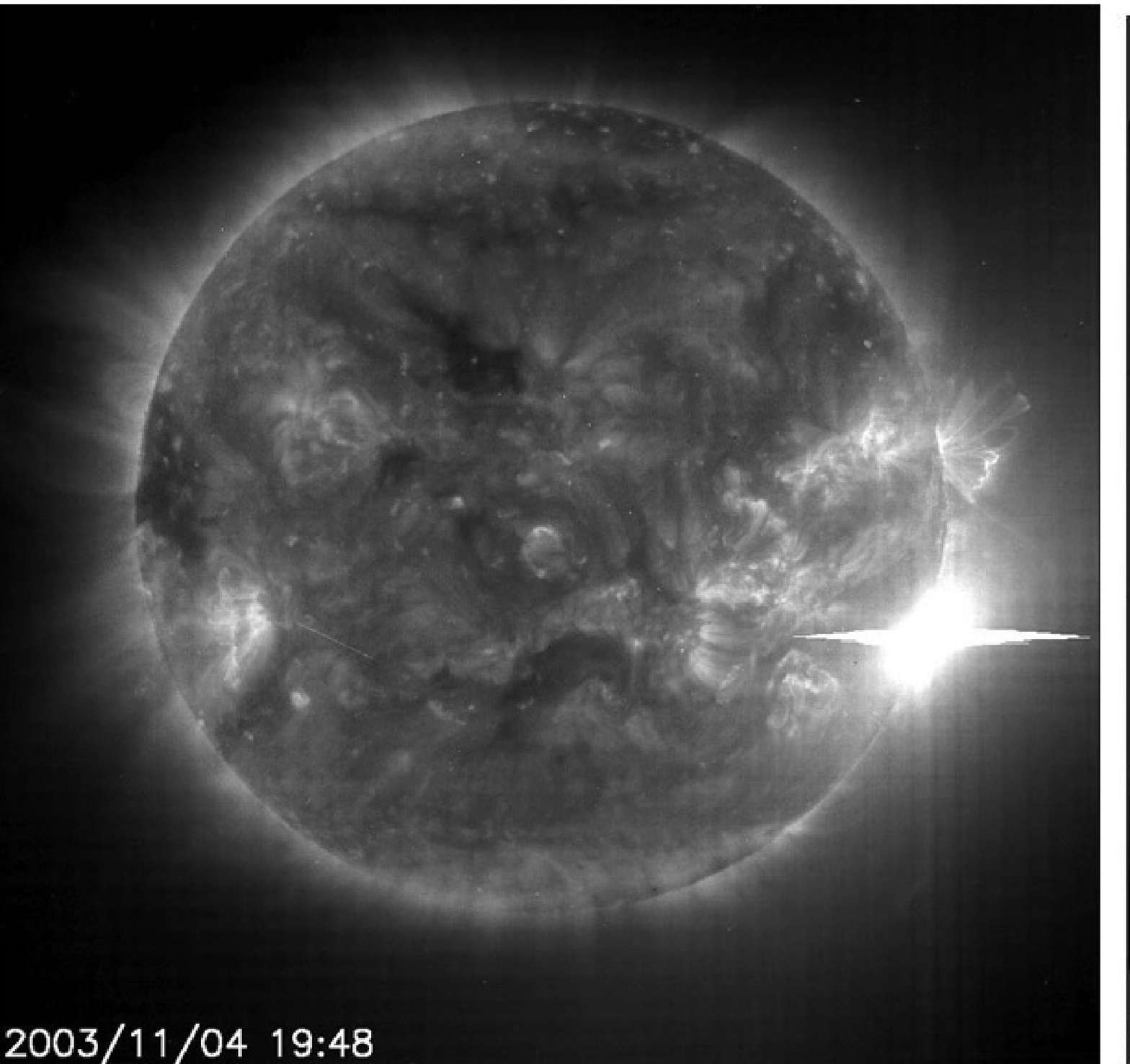,width=13cm}{X Solar Flare on
4th November 2003 observed by GOES and SOHO satellites: note the
sharp outburst at 19:48-20:08 UT.}

\section{Conclusions}
\label{sect:conclusion}
The recent solar flare occurred on October-November $2003$, as
large as the September $29th, 1989$ one, might be an exceptional
source of cosmic, gamma, neutron rays and neutrinos. Their
minimum event number at Super-Kamiokande
${N}_{{ev}_{\bar{\nu}_{\mu}}}\simeq {N}_{{ev}_{\bar{\nu}_{e}}}
       \simeq 2 (\frac{<{E}_{{\nu}_{\mu}}>}{200~MeV}(\frac{<{E}_{FL}>}{{10}^{31}~erg})$ ;
     ${N}_{{ev}_{{\nu}_{\mu}}}\simeq {N}_{{ev}_{{\nu}_{e}}} $
   is near or above unity. The background
due to  energetic atmospheric neutrinos at the Japanese detector
is nearly $5.8$ event a day corresponding to a rate
${\Gamma\simeq 6.7 10^{-5}} s^{-1}$. The lowest and highest
predicted event numbers ($1\div5$)~${\eta}$, (${\eta}\leq{1}$)
within the narrow time range
 defined by the sharp X burst ($100 s$), are above the background.  Indeed the probability to find by chance one
neutrino event within a $1-2$ minute ${\Delta}t \simeq 10^2 s$ in
that interval is $P\simeq\Gamma \cdot{{\Delta}T}\simeq 6.7
\cdot10^{-3}$. For a Poisson distribution the probability to find
$n=1,2, 3, 4, 5$ events in a  narrow time window might reach
extremely small values:
$
 {{P}_{n}}\cong\frac{P^{n}}{n!}=( 6.7
\cdot 10^{-3}, 2.25 \cdot 10^{-5},  5 \cdot 10^{-8}, 8.39
10^{-11}, 1.1\cdot10^{-13}). $
Therefore the possible presence of one or more high energetic
(tens-hundred MeVs) positrons (or better positive muons)
 as well as negative electrons or muons,  in Super-Kamiokande
 at X-flare onset time, may be a well defined signature of the solar
neutrino flare. A surprising discover of the complete mixing from
the $\tau$ appearance  may occur for hard (${E}_{{\nu}_{\mu}}
\rightarrow {E}_{{\nu}_{\tau}}\simeq> 4 GeV$) flare spectra. A
steep proton flare spectrum, where a large flare energy fraction
is at a low proton energies may reduce ${\sigma}_{pp}$ inelastic
cross-sections and increase the elastic ones, reducing the
pion-neutrino creations.  At low flare energy ${E}_{FL}<10^{32}$
erg, any neutrino muon spectra where
$\bar{E}_{\bar{\nu}_{\mu}}<$100 MeV, or any proton-magnetic field
interaction may suppress our estimates. Therefore our
considerations are only preliminary and they must be taken
cautiously (given the delicate chain of assumptions and
simplifications). We hope to stimulate with our work related
research in gamma/optical wavelengths, in the study of the neutron
component of cosmic rays, and in neutrino underground detectors to
investigate the solar activity. In particular we suggest to
control the very Super-Kamiokande data records on $October-
November$ solar flare X-radio peak activity, namely on
$26-28-30th$ October and $2nd-4th$ and $13$ November X-ray onset
(see figures below for time details). We like to  point the
attention to the hard X onset at $19:48$ U.T. on $4th$ November
$2003$. Finally we notice that the new larger neutrino detectors
such as UNO might be at the same time ideal laboratories for solar
neutrino flare and flavour mixing, as well as rapid alert system
monitoring coronal mass ejection dangerous for orbiting
satellites.

\acknowledgments The author wishes to thank Prof. M. Parisi,Prof.
M. Gaspero,  P.De Sanctis Lucentini, M. De Santis, Dr. Cristina
Leto and Dr. Marco Grossi for valuable suggestions.


\begin{thebibliography}{99}
\bibitem{ref0}{ F.Bachelet and A.M. Conforto, Nuovo
Cimento,3.1153(1956);  J.Simpson, Proc.National Acad. of Sc.of
USA,Vol43,42,(1957).}
\bibitem{ref1}{M. Alessio, L. Allegri, D.
Fargion, S. Improta, N. Iucci, M. Parisi, G.
Villoresi,N.L.Zangrilli, Il Nuovo Cimento, 14C, 53-60, (1991).}
\bibitem{ref2}{M. Dryer, Space Science Reviews, 15 (1974), 403-468.}
\bibitem{ref3}{V. S. Berezinsky, C. Castagnoli and P. Galeotti,Il Nuovo Cimento, 8C, 185 (1985).}%
\bibitem{ref4}{D.Fargion,Maxim Khlopov; Astroparticle Vol.19, 3,p.441-446,(2003)}
\bibitem{ref5}{M. E. Machado and J. L. Linsky, Solar Phys., 42, 395 (1975).}%
\bibitem{ref6}{A. Dar and S. P. Rosen, Preprint 27803 - Los Alamos Th. Div., August (1984).}%
\bibitem{ref7}{T. K. Gaisser, T. Stanev, G. Barr, Preprint Bartol Research Inst., 22/01/89, BA-88-1.}
\bibitem{ref8}{W. C. Haxton, Phys. Rev. D, 36, 2283, (1987).}
\bibitem{ref9}{S.N.Karpov,L.I.Miroshnichenko,E.V.Vashenyuk, Il Nuovo Cimento, 21C, 551, (1998)}
\bibitem{ref10}{D.Fargion; adsabs.harvard.edu/abs/1989STIN...9023331F; Preprint INFN n.$721$; $19Dec$.(1989)}
\bibitem{Fargion 2002}{ Fargion D, Ap. J. 570, 909-925;(2002) }
\bibitem{bemporad} {C.~Bemporad, G.~Gratta and P.~Vogel, Rev.\ Mod.\ Phys.\  {74}
297(2002)}
\bibitem{strumia} {Strumia A., Vissani F., astro-ph/0302055, (2003)}
 \bibitem{89} {L.I.Miroshnichenko, C.A.De Koning,R.
 Perez-Enriquez;SpaceScienceReviews 91; 615-715,(2000)}
 \bibitem{Lin} {R. P. Lin, et all.Ap.J, Volume 595, Number 2, Part 2;(2003)}.
 \bibitem{Bodek} {A.Bodek, H.Budd and J.Arrington hep-ex/0309024,
 (2003)}


\end{thebibliography}
\end{document}